\documentclass[11pt]{article}

\usepackage{amssymb, amsmath}
\usepackage{graphicx}
\usepackage{cite}

\newcommand{\cF}{\mathcal{F}}

\def\be{\begin{equation}}
\def\ee{\end{equation}}
\def\bea{\begin{eqnarray}}
\def\eea{\end{eqnarray}}

\title{ \bf{Electrovacuum Near-horizon Geometries in Four and Five Dimensions}}

\author{Hari K. Kunduri$^a$\footnote{hkunduri@phys.ualberta.ca }  \\ \\
 Theoretical Physics Institute, Department of Physics \\
University of Alberta, Edmonton, T6G 2J1, Canada
 }

\date{}

\begin{document}

\maketitle

\begin{picture}(0,0)(0,0)
\put(350, 225){Alberta-THY-07-11}
\end{picture}

\vskip1.5cm
\begin{abstract}
Associated to every stationary extremal black hole is a unique near-horizon geometry, itself a solution of the field equations. These latter spacetimes are more tractable to analyze and most importantly, retain properties of the original black hole which are intrinsic to the event horizon.  After reviewing general features of near-horizon geometries, such as $SO(2,1)$ symmetry enhancement,  I report on recent work on stationary, charged extremal black hole solutions of the Einstein-Maxwell equations with a negative cosmological constant in four dimensions and present a classification of near-horizon geometries of black holes on this kind. In five dimensions, charged extremal black hole solutions to minimal (gauged) supergravity, which arises naturally in string theory and the gauge theory/gravity correspondence, are considered. I consider the classification of near-horizon geometries for the subset of such black holes which are supersymmetric.  Recent progress on the classification problem in the general extremal, non-supersymmetric case is also discussed.

\end{abstract}

\newpage

\section{Introduction}
The classification of all asymptotically flat (for dimensions $D >4$) and asymptotically  Anti-de Sitter (AdS) (for $D \geq 4$) stationary black holes is a challenging open problem.  Two important recently established general results are that the event horizon of such black holes must be a Killing horizon \cite{HIW, IM} and that in the asymptotically flat case, spatial cross-sections of the event horizon must be of positive Yamabe type \cite{GS} (admit a metric of everywhere positive curvature). Focussing on \emph{extremal} black holes, characterised by vanishing surface gravity, allows us to study the difficult classification problem in a constrained setting.  Furthermore, the condition is sufficiently weak (e.g. vacuum Kerr can be extremal) that it captures a wide subset of the full space of solutions.
\par  Extremal black holes have also been essential for uncovering the origin of the black hole entropy from a statistical counting of their microstates. This was first performed within string theory for the special class of \emph{supersymmetric}, asymptotically flat black holes.  Such black holes are necessarily extremal, and recent work strongly suggests that it is this property that is responsible for the successful calculation of the entropy.  Strong evidence to support this claim is provided by the exact microstate counting of several extremal, non-supersymmetric black holes.
\par  It is therefore desirable to develop techniques to classify extremal black holes. A useful approach is to employ the fact that every extremal black hole admits a near-horizon limit that yields a spacetime that solves the same theory.  The resulting near-horizon geometry retains properties of the black hole intrinsic to the event horizon.   As we explain below, the classification of these geometries is equivalent to a more tractable $(D-2)$ -dimensional problem on a closed Riemannian manifold. The classification allows us to deduce general statements on the full space of extremal solutions in a given theory.  Combined with additional global information on the spacetime,  this can provide a method to solve the uniqueness/classification problem for extremal black holes (e.g. \cite{CRT2, R}). Further, the existence of enhancement of symmetry in the near-horizon region \cite{KLR2} appears to play a key role within the quantum description.
\par This brief article focusses on the near-horizon geometries of extremal black holes with non-vanishing Maxwell fields. After a brief review of properties of near-horizon geometries, the near-horizon symmetry enhancement that emerges dynamically in a wide class of theories for $D =4 \, 5$ is discussed. Next, the classification problem in $D=4$ Einstein-Maxwell theory with a negative cosmological constant, and in $D=5$ for (gauged) minimal supergravity is considered.  We conclude with a short summary.

\section{Extremal Black holes and Near-horizon Geometries}
\subsection{The near-horizon limit}
Consider a stationary black hole spacetime in $D$ dimensions\footnote{In fact many of the results of the section will hold for any spacetime containing a degenerate Killing horizon, but we will be concerned with the black hole case}.  We will focus on black holes which are asymptotically flat or asymptotically Anti-de Sitter. The event horizon $\mathcal{N}$ of a \emph{static} non-extremal black hole is a Killing horizon with respect to the Killing field which is timelike at spatial infinity. We will assume that this result also holds for degenerate (extremal) static black holes. For a  non-static, rotating non-extremal black hole, the event horizon is also a Killing horizon; in particular, it has been proved such spacetimes must be \emph{axisymmetric}, i.e. admit an additional spacelike Killing field with closed orbits\cite{HIW, IM}.  The proof of this statement for extremal black holes has recently been given in four dimensions and there are partial results in five dimensions\cite{HI}. We will assume throughout this article that the event horizon of a stationary, extremal black hole in five and higher dimensions is also a Killing horizon with respect to some Killing field $V$.  In a neighbourhood of the future event horizon $\mathcal{N}^+$ of an extremal black hole, one may always introduce Gaussian null coordinates $(v,r,x^a)$ such that the metric takes the form
\begin{equation}\label{GNCnonext}
ds^2 = r^2 F(r,x) dv^2 + 2 dv dr + 2 r h_a (r,x) dv dx^a + \gamma_{ab}(r,x) dx^a dx^b.
\end{equation} The Killing field $V = \partial/ \partial v$ is null at $r=0$ and $\partial/\partial r$ is tangent to null geodesics that point `out' of $\mathcal{N}^+$.  The function $F$, one form $h$, and symmetric rank two tensor $\gamma$ are smooth fields on spacetime.  The label $x$ refers collectively to coordinates $x^a$ which are a local chart on spatial cross-sections of the horizon $\mathcal{H}$ which is a closed (compact without boundary) $D-2$ dimensional manifold with Riemannian metric $\gamma_{ab}(0,x)$.   An extremal black hole is characterized by a degenerate event horizon, i.e. with vanishing surface gravity $\kappa=0$, which is reflected in (\ref{GNCnonext}) by $V^2 = O(r^2)$.  We may then take the \emph{near-horizon} limit of (\ref{GNCnonext})
\begin{equation}\label{limit}
v \to \frac{v}{\epsilon}, \qquad r \to \epsilon r, \qquad \textrm{with}\phantom{a}\epsilon \to 0 .
\end{equation} resulting in a near-horizon geometry described by the metric $g_{NH}$ with line element
\begin{equation}\label{NHG}
ds^2 = r^2 F(x) dv^2 + 2dvdr + 2 r h_a(x) dv dx^a + \gamma_{ab}dx^a dx^b.
\end{equation} Note that this limit is well-defined only for an extremal black hole, or more generally, a spacetime containing a degenerate Killing horizon.  The geometry is fully specified by the fields $(F, h_a, \gamma_{ab})$ which are respectively a scalar, one-form, and metric defined purely on $\mathcal{H}$ .  A schematic spacetime diagram illustrating the near-horizon geometry is given in Figure 1.
 \begin{figure}[h] \label{fig1}
   \centering
   \includegraphics[scale=0.3]{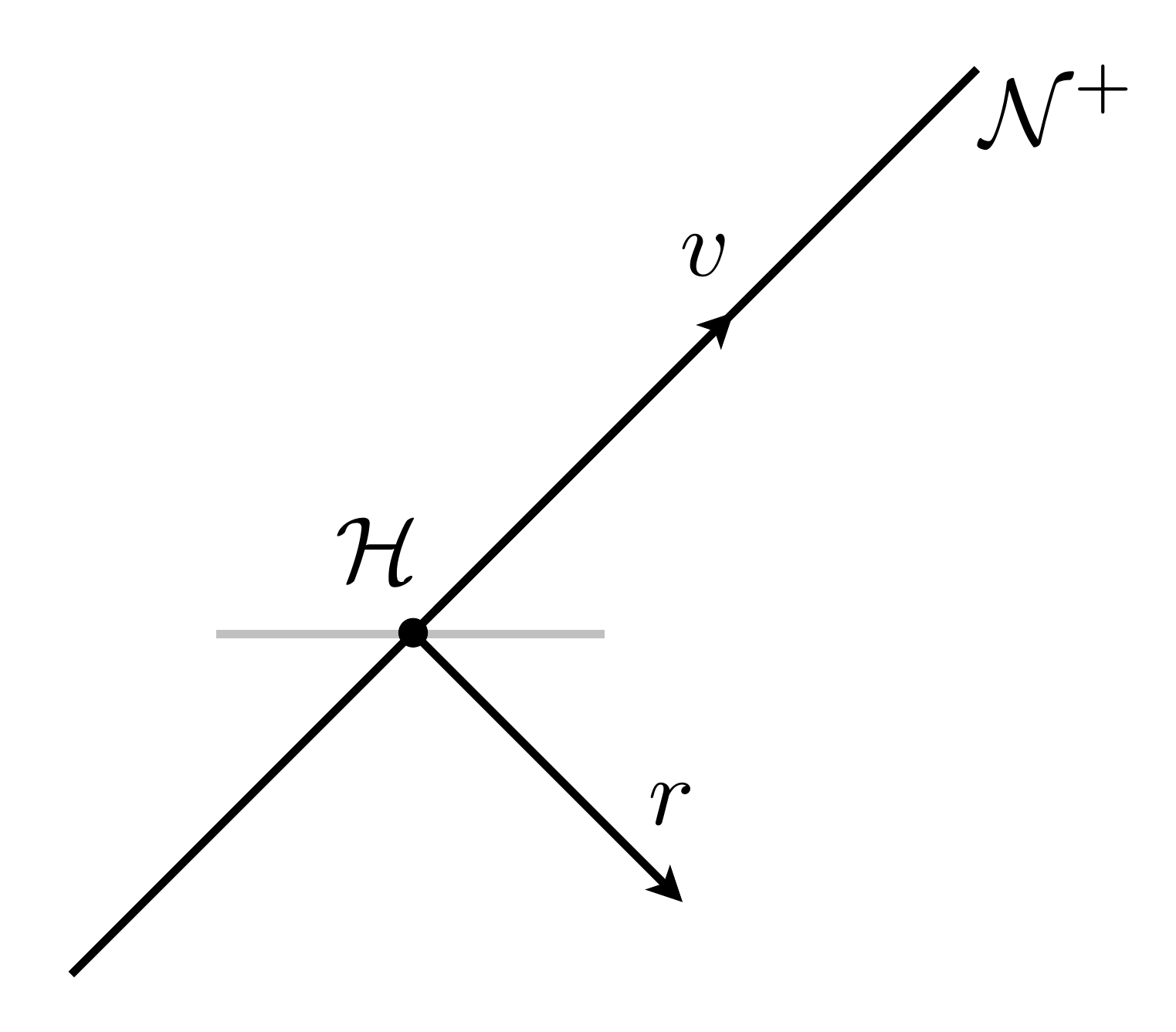}
   \caption{spacetime diagram of near-horizon region}
  \end{figure}
The near-horizon limit must also be applied to the other matter fields that describe the full black hole solution.  Under the limit, the scalar fields become smooth functions on $\mathcal{H}$.  The Maxwell field in general does \emph{not} possess a near-horizon limit, but it will for solutions of the theories we will consider as a consequence of the well known result $R_{\mu \nu} V^\mu V^\nu |_\mathcal{N} = 0$.  Using this and the Bianchi identity $dF=0$, one can show that, assuming analyticity, the field strength in the near-horizon limit must take the form
\begin{eqnarray}\label{maxNH}
 \cF_{NH} = d(\Delta(x) r dv) + \hat{\mathcal{F}},
\end{eqnarray} where  $\Delta(x) = -\cF_{vr}$ is a smooth function and $\hat{\mathcal{F}}$ is a closed 2-form defined on $\mathcal{H}$.  \par It is important to emphasize that the near-horizon limit~(\ref{limit}) maps solutions of a given theory to solutions of the same theory.  Hence the set of fields $(g_{NH}, \cF_{NH}, \phi)$ constitute a solution in its own right, which contains key information regarding the `parent' extremal black hole, in particular the geometry of $\mathcal{H}$.   Before turning to the classification of such spacetimes,  we first review a useful result concerning the symmetries of near-horizon geometries with rotational symmetries.

\subsection{Near-Horizon Symmetry Enhancement}
The metric~(\ref{GNCnonext}) possesses a two-dimensional (2d) non-Abelian group $G_2$ generated by $\partial / \partial v$ and $v \partial/\partial v + r \partial / \partial r$. The orbits are 2d for $r\neq 0$ and 1d when $r=0$. Remarkably, the isometry group $G_2$ is enhanced dynamically to $O(2,1)$ as a result of Einstein's equations~\cite{KLR2}. For \emph{static} black holes, this can in fact be seen to arise kinematically. The Killing field $V$ must be hypersurface orthogonal, $V \wedge dV = 0$ and hence the associated near-horizon geometry is also static. This occurs if, and only if $dh = 0$ and $dF = Fh$. By a suitable coordinate transformation~\cite{KLR2}, one can show the near-horizon metric must be locally a warped product of a 2d maximally symmetric space with non-positive curvature ($AdS_2$ or $\mathbb{R}^{1,1}$) with a compact manifold $\mathcal{H}$. Hence $G_2$ is enhanced to a local \footnote{the symmetry is global if $\mathcal{H}$ is simply connected.} $O(2,1)$ in the $AdS_2$ case or a local Poincare symmetry in the Minkowski case. The Maxwell field, if present, is  only invariant under $SO(2,1)$. A typical example of a spacetime of this type is the near-horizon geometry of the extremal Reissner-Nordstrom solution, $AdS_2 \times S^2$.
\par Next consider stationary extremal \emph{rotating} black holes, for which the stationary Killing field at infinity is not null on $\mathcal{N}$. As mentioned above, in four dimensions such solutions have been proved to be axisymmetric. As a consequence, the near-horizon is cohomogeneity-1. In five and higher dimensions, assuming that the rigidity results~\cite{HIW} extend to the extremal case, stationary rotating black holes also admit an additional rotational symmetry, so that their isometry group must be at least $\mathbb{R}\times U(1)$.  Hence their associated near-horizon geometries will inherit a $G_2 \times U(1)$ isometry. All \emph{known} 5d solutions in fact admit an additional $U(1)$ isometry, which, as in four dimensions, renders their near-horizon geometries cohomogeneity-1. Let us focus on black holes with $U(1)^{D-3}$ rotational symmetries. For $D>5$ such solutions cannot be asymptotically flat or Anti-de Sitter, because the rotational group $SO(D-1)$ does not admit a $U(1)^{D-3}$ subgroup. However, they can be naturally interpreted as black branes, which upon dimensional reduction, describe black holes in $D=4$ or $D=5$. Note that the existence of a global $U(1)^{D-3}$ action imposes some restrictions on the topology of $\mathcal{H}$\cite{Chr}: for $D=4$, $\mathcal{H}$ must have topology $S^2$ or $T^2$, whereas for $D=5$, $\mathcal{H}$ must have topology $S^3$ (or quotients), $S^1 \times S^2$, or $T^3$.
\par Consider a general second-order theory of gravity in $D=4,5$ coupled to uncharged scalars $\phi^A$ ($A=1...M$) and Maxwell fields $\cF^I$, $I=(1...N)$ with action
\begin{eqnarray}\label{gaction}
S &=& \int (R - V(\phi))\star{1} - \frac{1}{2}f_{AB}(\phi)\star d{\phi}^A \wedge d\phi^B  - \frac{1}{2}g_{IJ}(\phi) \star \cF^I \wedge \cF^J + S_{\textrm{top}} \,, \\
S_{\textrm{top}} &=& \left\{
\begin{array}{l l}
  \frac{1}{2} \int h_{IJ}(\phi) \cF^I \wedge \cF^J & \quad \mbox{$D=4$}\\
  \frac{1}{3!} \int C_{IJK} \cF^I \wedge \cF^J \wedge A^K  & \quad \mbox{$D=5$}\\ \end{array} \right. \,.\nonumber
\end{eqnarray}  Here $V(\phi)$ is an arbitrary scalar potential and the $C_{IJK}$ are constants.  Many theories of interest are contained within~(\ref{gaction}), such as vacuum gravity including a cosmological constant, Einstein-Maxwell theory,  and several (gauged) supergravities arising from compactifications from string theory.  We may then prove the following: \cite{KLR2}
\paragraph{Theorem} Consider a stationary, asymptotically flat or asymptotically Anti-de Sitter extremal black hole solution of (\ref{gaction}) with $U(1)^{D-3}$ rotational symmetries.  Then its near-horizon geometry has an enhanced $G_3 \times U(1)^{D-3}$ isometry group, where $G_3 = SO(2,1)$ or the orientation-preserving subgroup of the 2d Poincar\'e group. The latter is excluded if $g_{IJ}, f_{AB}$ are positive definite, $V(\phi) \leq 0$, and spatial cross sections of the event horizon $\mathcal{H}$ are non-toroidal.  \\
\par \noindent  This result is proved \cite{KLR2} by integrating certain components of the field equations arising from~(\ref{gaction}) which imply, upon suitable rescaling of the radial coordinate defined in (\ref{NHG}), that the near-horizon solution must take the form
\begin{eqnarray}
\label{so21met} ds^2 &=& \Gamma(\rho)\left[A_0 r^2 dv^2 + 2dv dr\right] + d\rho^2 + \gamma_{ij}(\rho)\left( dx^i + k^i r dv\right)\left(dx^j + k^j r dv\right) \\
\cF_{NH} ^I &=& -\Gamma(\rho) \Delta^I(\rho) dv \wedge dr + \cF(\rho)_{\rho i}^I d\rho \wedge \left(dx^i + k^i r dv\right),  \quad \phi^A = \phi^A(\rho) \label{so21max}
\end{eqnarray} where $A_0, k^i$ are constants, $\partial / \partial x^i$ are Killing fields generating the rotational symmetries ($i=1..D-3$) and $\Gamma(\rho) > 0, \, \gamma_{ij}(\rho)$ are smooth functions of the horizon coordinate $\rho$ which parameterizes the orbit space $\mathcal{H}/U(1)^{D-3}$. If $A_0 < 0$, the term in the square brackets of (\ref{so21met}) is simply $AdS_2$. The case $A_0 \geq 0$ is excluded subject to the extra conditions listed in the Theorem.  Hence the total isometry group of the near-horizon geometry is $SO(2,1) \times U(1)^{D-3}$, where the $SO(2,1)$ has 3d orbits if $k^i \neq 0$ and 2d if $k^i = 0$.  The near-horizon geometry is a fibration of $\mathcal{H}$ over $AdS_2$.   In general, the metric~(\ref{so21met}) is non-static for $k^i \neq 0$, but in the special case $\gamma_{ij}k^i k^j = -A_0 \Gamma$ and $\gamma_{ij} k^j = c_i \Gamma$, where $c_i$ are constant, then the spacetime is static, the symmetry is further enhanced to a local $O(2,2) \times U(1)^{D-4}$, and the near-horizon geometry is a warped product of $AdS_3$ and a closed $D-3$-dimensional manifold $\hat{\mathcal{H}}$. In this case, $\mathcal{H} \sim S^1 \times \hat{\mathcal{H}}$.
\section{Classifications of Electrovacuum Near-horizon geometries}
We now turn to the classification of near-horizon geometries for special cases of interest of the class of theories described by (\ref{gaction}). Determining all near-horizon geometries of a given theory yields both the allowed topologies and geometries of spatial cross sections of extremal black hole solutions. In particular, we can use classifications to \emph{rule out} the existence of certain topologies. However, the converse statement is not necessarily true; the existence of a near-horizon geometry does not guarantee the existence of an associated full extremal black hole solution. The general strategy employed is to reduce the full spacetime equations to a set of covariant equations on $\mathcal{H}$, solve for all possible \emph{near-horizon data} $(F,h_a, \gamma_{ab},  \Delta, \hat{\mathcal{F}})$, and finally impose regularity.  We may also use the near-horizon data to compute certain conserved charges. We will focus on solutions with non-vanishing Maxwell fields. The classification problem in the pure vacuum case (including a cosmological constant $\Lambda $) has been solved in the static case in all dimensions\cite{CRT1} and in the non-static case, a classification of near-horizon geometries with $U(1)^{D-3}$ rotational symmetries for $D=4,\, 5$ was achieved in~\cite{KLvac}.  This has been subsequently extended to $D \geq 6$ for $\Lambda =0$ \cite{HIKK}.
\subsection{Four dimensions} Consider Einstein-Maxwell theory with a negative cosmological constant. We wish to find all possible near-horizon metrics and Maxwell fields of the form (\ref{NHG}) and (\ref{maxNH}) satisfying the coupled Einstein-Maxwell field equations. A lengthy calculation\footnote{These equations for $\Lambda =0$ were first analysed in \cite{Hac} and subsequently in \cite{LP} in the context of extremal isolated horizons} \cite{KL4d} shows this is equivalent to the following covariant equations on $\mathcal{H}$.
\begin{eqnarray}
 \mathcal{R}_{ab} &=& \frac{1}{2}h_a h_b - \nabla_{(a}h_{b)} + \Lambda \gamma_{ab} + 2\hat{\cF}_{ac}\hat{\cF}_{bd}\gamma^{cd} + \Delta^2\gamma_{ab} - \frac{\gamma_{ab}}{2}\hat{\cF}^2 \label{4dmaxwell} \\
F &=& \frac{1}{2}h_ah^a - \frac{1}{2}\nabla_a h^a + \Lambda -\Delta^2 - \frac{\hat{\cF}^2}{2}   \qquad d\star_2 \hat{\cF} = \star_2 i_h\hat{\cF}+ \star_2 (d\Delta-\Delta h). \nonumber
\eea where $\mathcal{R}_{ab}$, $\nabla$ and $\star_2$ are the Ricci tensor, the covariant derivative and Hodge dual of the 2d horizon metric $\gamma_{ab}$.  \par For static near-horizon geometries, the condition that $V$ be hypersurface-orthogonal,  $V \wedge dV = 0$, requires $dh = 0, dF = Fh$ and $d\Delta = h\Delta$. It turns out \cite{KL4d} these conditions are restrictive enough to classify all possible regular solutions without further assumptions.  Introducing the globally defined function $\phi = \star_2 \hat{\mathcal{F}}$, we find $h = 0$, and $F, \Delta$ and $\phi$ must be constants satisfying $F = -\Delta^2 - \phi^2 + \Lambda \leq 0$ (since $\Lambda \leq 0$ and $\mathcal{F} \neq 0$). The full near horizon geometry is a direct product  $AdS_2 \times \mathcal{H}$ where $\mathcal{H}$ is a 2d closed manifold equipped with an Einstein metric $\gamma_{ab}$ satisfying
\be \mathcal{R}_{ab}= \lambda \gamma_{ab}, \qquad \lambda \equiv \Delta^2+\phi^2+\Lambda \ee which implies $\gamma_{ab}$ is locally
isometric to one of the maximally symmetric metrics on $S^2,T^2,H^2$
depending on the sign of $\lambda$.  As we are mainly interested in asymptotically flat and globally AdS$_4$ black holes, topological censorship permits only $\mathcal{H} = S^2$, the near-horizon limit of Reissner-Nordstrom-(AdS$_4$). In the asymptotically locally AdS$_4$ case, the only possible static near-horizon geometries must have $\mathcal{H} = \Sigma_g$ (a higher genus surface) with $\lambda <0$ or $T^2$  $\lambda = 0$. The classification of static eletrovacuum, asymptotically black holes is given in \cite{CT}.
\par The non-static case is made tractable by noting that a 4d stationary rotating\footnote{there could be a non-static, non-rotating black hole which need not be axisymmetric, although this cannot occur at least in the asymptotically flat case\cite{SW}.} asymptotically flat or globally AdS$_4$ extremal black hole must be axisymmetric\cite{HI}.  The near-horizon geometry will inherit $m$, which leaves the near-horizon data invariant. Hence from the Theorem in the previous section (or directly from~(\ref{4dmaxwell})) we deduce that the near-horizon metric and Maxwell field must take the form~(\ref{so21met}) and (\ref{so21max}) respectively. This makes it possible to then integrate the remaining second-order equations (\ref{4dmaxwell}) for the near-horizon data to prove that the \emph{unique} non-static and axisymmetric near-horizon geometry with a compact horizon section of $S^2$ topology is given by \cite{Hac, LP, KL4d}:
\bea\label{NSsolEM4d}
ds^2&=& \Gamma[ -C^2 r^2 dv^2 +2dvdr] + \frac{\Gamma d\sigma^2}{Q} + \frac{Q}{\Gamma} (dx+rdv)^2 \\
\cF &=& d[ E (r dv + dx)] , \qquad   E=  \frac{ \sigma e\cos\alpha - \left(
\beta^{-1}- \frac{\beta \sigma^2}{4} \right) e\sin \alpha}{ \Gamma} \nonumber
\eea where \be\label{NSdata} \Gamma =
\beta^{-1}+ \frac{\beta\sigma^2}{4}, \qquad Q = -\frac{\beta
\Lambda}{12} \sigma^4 - (C^2+2\Lambda
\beta^{-1})\sigma^2+4\beta^{-3}(C^2\beta +\Lambda -e^2\beta^2) \ee
and $C,\beta,e>0$ and $\alpha$ are constants.  A scaling symmetry allows one to fix one of these to any desired value. Therefore, it is a 3-parameter family.  The polynomial $Q(\sigma)$ must have a roots at $\pm \sigma_2$ and the coordinate ranges are $-\sigma_2 \leq \sigma \leq \sigma_2$ (with $Q>0$ inside this interval) and $x$ is periodically identified in such a way to remove the conical singularities at $\sigma=\pm\sigma_2$. It is straightforward to show that (\ref{NSsolEM4d}) is globally isometric to the near-horizon limit of the Kerr-Newman-(AdS$_4$) black hole \cite{KL4d}. The proves that the near-horizon geometry of \emph{any} stationary, extremal rotating asymptotically globally AdS$_4$ black hole must be given by that of Kerr-Newman-AdS$_4$. An important application of these results in the $\Lambda = 0$ case is the recent proof that the extremal Kerr-Newman solution is the unique asymptotically flat rotating degenerate electrovacuum black hole \cite{CN}.  \par
Finally, recall that Einstein-Maxwell-$\Lambda$ theory is the bosonic sector of $N=2$ minimal gauged supergravity with gauge coupling $g$ defined by $\Lambda = -3g^2$.  Supersymmetric black holes are necessarily extremal and hence the near-horizon geometry of any supersymmetric black hole in the theory is automatically contained within our classification.  We can identify the subset of solutions which are supersymmetric by inspecting the integrability conditions for the existence of a Killing spinor.  In the static case, only the $\lambda <0$ case with $\Delta^2 + \phi^2 = g^2$ is allowed. In the non-static case, the most general supersymmetric near-horizon geometry is given by (\ref{NSsolEM4d}) subject to the additional conditions $\cos\alpha = 0$ and $\beta e^2 = C^2 - 4 g^2$.  Each of these solutions preserves half the supersymmetry.  From these results it follows that the near-horizon geometry of \emph{any} supersymmetric AdS$_4$ rotating black hole is that of the most general known solution. In the ungauged theory ($g=0$) the near-horizon analysis has been used to prove (under the additional hypothesis that there are no null orbits of the stationary Killing vector within the domain of outer communications) that the extremal Reissner-Nordstrom solution is the unique asymptotically flat supersymmetric black hole with connected horizon\cite{CRT2}.

\subsection{Five Dimensions} We will restrict our attention to the classification of near-horizon geometries in minimal ungauged and gauged 5d supergravity, which admit asymptotically flat and asymptotically globally AdS$_5$ black holes respectively. These theories are described by the action
\begin{equation}\label{5daction}
S = \frac{1}{16\pi G_5}\int \star 1 \left(R + \frac{12}{\ell^2}\right) - 2 \cF \wedge \star \cF - \frac{8}{3\sqrt{3}} \cF \wedge \cF \wedge A
\end{equation} where $\ell$ denotes the AdS$_5$ length scale. For the ungauged theory, this cosmological term is absent. Note that the ungauged theory~(\ref{5daction}) arises naturally as a consistent truncation of 11d supergravity on $T^6$, whereas the gauged theory arises from Type IIB supergravity reduced on $S^5$ and is relevant to studies of the AdS/CFT correspondence.  The spacetime field equations are equivalent to
\begin{eqnarray}
 \mathcal{R}_{ab} &=& \frac{1}{2}h_a h_b - \nabla_{(a}h_{b)}  -\frac{4}{\ell^2}\gamma_{ab} + 2\hat{\cF}_{ac}\hat{\cF}_{bd}\gamma^{cd} + \frac{1}{2}\Delta^2\gamma_{ab} - \frac{\gamma_{ab}}{3}\hat{\cF}^2 \label{5dRiceq} \\
F &=& \frac{1}{2}h_ah^a - \frac{1}{2}\nabla_a h^a -\frac{4}{\ell^2} -\Delta^2 - \frac{\hat{\cF}^2}{3} \label{Feq} ,
\label{maxeq} \qquad d\star_3 \hat{\cF} = -\star_3 i_h\hat{\cF}- \frac{\sqrt{3}}{2} \star_3 (d\Delta-\Delta h)+2 \Delta \hat{\cF} \nonumber
\end{eqnarray} where $R_{ab}$, $\nabla$ and $\star_3$ are the Ricci tensor, the covariant derivative and Hodge dual of the 3d metric $\gamma_{ab}$. Note that for convenience we have made a constant rescaling of $\Delta$ defined in (\ref{maxNH}). The main technical obstacle in extending the 4d results described above is that $\mathcal{H}$ is 3d and there are more independent, coupled components in $\gamma_{ab}$ and $\hat{\mathcal{F}}$, which makes it more difficult to solve for the near-horizon data. Moreover, the electric and magnetic parts of the Maxwell field are no longer related by duality transformations. To overcome these difficulties, it proves useful to impose additional global constraints. We describe the results in detail below.
\subsection{Near-horizon geometries of Supersymmetric Black holes}
{\bf \emph{Asymptotically Flat Black Holes -}} Supersymmetric black holes in five dimensions have played a crucial role in understanding the microscopic origin of the Bekenstein-Hawking entropy within string theory.  An important first step towards mapping out the space of such solutions was the classification of near-horizon geometries of supersymmetric black holes in \emph{ungauged} minimal supergravity achieved in~\cite{R}. In terms of the near-horizon data, supersymmetry is sufficient to show that $h$ must be a Killing field and $\Delta$ and $F$ must be constant. This allows one to completely solve~(\ref{5dRiceq}) for all possible regular near-horizon geometries with closed $\mathcal{H}$. The only allowed possibilities are \emph{homogenous} spacetimes: (i) $AdS_3 \times S^2$, with $\mathcal{H} = S^1 \times S^2$; (ii) an $S^3$ fibration over AdS$_2$ with $\mathcal{H}$ a homogenously squashed $S^3$ and (iii) $\mathbb{R}^{1,1}\times T^3$ with $\mathcal{H} = T^3$ with its flat metric. In light of topological censorship, (iii) may be excluded. Cases (i) and (ii) have been explicitly realized as the near-horizon limits of the supersymmetric black ring \cite{susyring} and the BMPV black hole \cite{BMPV} respectively. An important application of this analysis is the proof that the BMPV black hole is the \emph{unique} supersymmetric black hole with $S^3$ horizon topology \cite{R}.\\ \par \noindent
{\bf \emph{Supersymmetric AdS$_5$ Black Holes -}}  It remains a challenging open problem to reproduce the entropy of these solutions using the AdS/CFT correspondence. Such black holes are expected to be dual to high-energy states in the strongly coupled $SU(N)$ conformal field theory on $\mathbb{R}\times S^3$. However, the black holes preserve only 2 supercharges (they are 1/16th BPS) and the corresponding supermultiplets may not be fully protected from quantum corrections. This is consistent with a computation of the degeneracy of these states in the free theory, and it is expected weak-coupling corrections must be taken into account \cite{MinMal}.  In the gravitational sector, a full classification of allowed black holes remains to be achieved.  The most general \emph{known} solutions have $\mathbb{R} \times U(1)^2$ symmetry, $S^3$ horizon topology, and possess four independent conserved charges \cite{KLRAdSBH}. However, the quantum states are expected to carry five such charges. A possible resolution to this apparent mismatch is that more general solutions exist (e.g. black rings). A systematic approach to search for such solutions is to classify the possible near-horizon geometries permitted in the theory, following the strategy employed in the ungauged case~\cite{R}. We will consider here only the minimal theory~(\ref{5daction}), but analogous results have been found in the more general theory containing additional vector multiplets \cite{Kunduri:2007qy}.  The inclusion of the cosmological constant term significantly complicates the problem. The first general near-horizon analysis was given in~\cite{GR1}, where a partial classification was obtained, focussing on $\mathcal{H}$ equipped with homogeneous metrics. Here, we restrict ourselves to near-horizon geometries with $U(1)^2$ symmetry\cite{KLRnoring}. The most general supersymmetric non-static near-horizon geometry of this kind is of the form~(\ref{NHG}) with \cite{KLRnoring}
\bea
&& \gamma_{ab}dx^a dx^b = \frac{\ell^2 \Gamma d\Gamma^2}{4 P(\Gamma)} + \left( C^2 \Gamma - \frac{\Delta_0^2}{\Gamma^2} \right) \left( dx^1 + \frac{\Delta_0  (\alpha_0 - \Gamma)}{C^2 \Gamma^3 - \Delta_0^2} dx^2 \right)^2 + \frac{4 \Gamma P(\Gamma)}{\ell^2 (C^2 \Gamma^3 - \Delta_0^2)} (dx^2)^2, \nonumber \\
  F&=& -\frac{\Delta_0}{\Gamma^2}, \qquad h = \Gamma^{-1} (k - d\Gamma), \qquad k = \frac{\partial}{\partial x^1}, \qquad P(\Gamma) = \Gamma^3 - \frac{C^2 \ell^2}{4} \left( \Gamma- \alpha_0 \right)^2 - \frac{\Delta_0^2}{C^2}
\eea with $C$ and $\Delta_0$ positive constants and $\alpha_0$ an arbitrary constant. One of these may be removed by a rescaling, leaving a 2-parameter set of solutions. We have $\Gamma_0 < \Gamma < \Gamma_1$ with $P(\Gamma) > 0$ in this interval and $\Gamma_0,\Gamma_1$ are roots of $P(\Gamma)$. Suitable linear combinations $(m_1,m_2)$ of the $\partial/\partial x^i$, which generate the $U(1)^2$ isometries, can be chosen so that $\gamma_{ab}$ smoothly extends to a cohomogeneity-1 metric on $\mathcal{H}$ with $S^3$ topology. This spacetime corresponds to the near-horizon geometry of the most general known supersymmetric AdS$_5$ black hole solution \cite{KLRAdSBH} of (\ref{5daction}). This suggests that the solution \cite{KLRAdSBH} may in fact be unique within its symmetry class. Interestingly, a local solution corresponding to the near-horizon of a supersymmetric AdS$_5$ \emph{black ring} was found in\cite{KLRnoring}, with a near-horizon that was a warped product of AdS$_3$ with a squashed $S^2$. However, the $S^2$ suffered from a conical singularity at one pole, implying that angular momenta and charge are insufficient to prevent the ring from self-collapse.

\subsection{Near-horizon geometries of non-supersymmetric extremal Black Holes} Supersymmetry effectively reduces (\ref{5dRiceq}) to a set of first order equations, greatly simplifying the classification problem. Non-supersymmetric near-horizon geometries may be thought as solution interpolating between vacuum~\cite{KLvac} and supersymmetric solutions~\cite{R}. Progress on the general extremal case in the ungauged theory has been made in certain cases.  Before discussing these results, it is worth introducing invariant notions of the electric and magnetic fields. Define the \emph{electric} field $\mathcal{E} = -i_V \cF$ and \emph{magnetic} field $\mathcal{B} = i_V \star \cF$ so that $\cF = -V^{-2} [ V \wedge \mathcal{E} + \star(V \wedge \mathcal{B})]$.  In terms of the near-horizon data ($F, h_a, \gamma_{ab}, \Delta, \hat{\cF}$) we have
\begin{equation}
\mathcal{E} = \frac{\sqrt{3}}{2} d ( r\Delta), \qquad \mathcal{B} = d(r\star_3 \hat{\cF}) - 2 r \Delta \hat{\cF} .
\end{equation}  This allows us to distinguish between solutions with non-vanishing electric fields ($\Delta \neq 0$) and/or magnetic fields ($\hat{\cF} \neq 0$). \par We focus first on \emph{static} near-horizon geometries \cite{KL5d}. Unlike pure Einstein-Maxwell theory, the action~(\ref{5daction}) admits non-static, non-rotating solutions (e.g. BMPV \cite{BMPV}), and moreover static near-horizons can arise as limits of both static and non-static extremal black holes\cite{susyring}.  Recall that we must have $dh = 0$, $dF = Fh$ and $d\Delta = h \Delta$. We will restrict attention to solutions with $U(1)^2$ rotational symmetry (see \cite{KL5d} for general results obtained without this assumption).  The starting point is to note that since $\mathcal{H}$ is closed, Hodge's theorem allows us to write $h = \beta + d\lambda$ where $\beta$ is a globally defined harmonic one-form and $\lambda$ a globally defined function.  The analysis splits into two cases, depending on whether $\beta =0$, which must be the case if $H^1(\mathcal{H}) = 0$, or $\beta \neq 0$. In the latter case, we can show that $F = \mathcal{E} =0$ and the \emph{only} regular near-horizon is a direct product of (a quotient of) AdS$_3$ and a round $S^2$ with $\mathcal{H} = S^1 \times S^2$. \par The case $\beta \neq 0$ results in near-horizon geometries which are warped products of AdS$_2$ with $\mathcal{H}$. If $\mathcal{B}=0$, then the only two possibilities are $\mathcal{H}$ = $S^3$ with its round metric and another solution in which $\mathcal{H}$ is an $S^3$ equipped with an inhomogeneous metric; while the former is clearly the near-horizon limit of extremal Reissner-Nordstrom,  it is not clear whether the latter corresponds to the near-horizon of an asymptotically flat extremal black hole. On the other hand, if $\mathcal{E}=0$, then we can only reduce all of~(\ref{5dRiceq}) to a single fourth-order ODE for one function. We have been able to prove that, should solutions exist, then $\mathcal{H} = S^1 \times S^2$ where the $S^2$ generically has an inhomogeneous metric. It is worth noting that in pure Einstein-Maxwell theory, a solution of precisely this kind exists: AdS$_2 \times S^1 \times S^2$, which is the near-horizon limit of the 4d dyonic Reissner-Nordstrom solution times a $S^1$. Finally, we turn to the most complicated situation, in which \emph{both} electric and magnetic fields are present. The classification reduces to a single third-order non-linear ODE. We have been unable to solve this generally.
The difficulty in determining all possible static near-horizon solutions of ungauged supergravity~(\ref{5daction}), even with the assumption of two rotational symmetries, suggests that the classification problem of extremal, asymptotically flat static black holes (in particular for non-vanishing magnetic fields) will be a formidable task.
\par Finally, we will consider recent work on the \emph{non-static} case currently under investigation\cite{KL2011}. The corresponding extremal, asymptotically flat black holes are expected to carry four conserved charges corresponding to mass, charge, and two independent angular momenta, as well as further non-conserved parameters such as `dipole charges' arising from non-trivial magnetic fields. Although extremality will impose one constraint on these parameters, the solution space will still be significantly larger than in four dimensions. Another difficultly is that solving (\ref{5dRiceq}) generally will lead to near-horizon geometries of black holes which are \emph{not} asymptotically flat  . \par  Nonetheless, for near-horizon geometries with $U(1)^2$ isometry, it \emph{is} possible to fully integrate for the near-horizon data. The strategy is to exploit the fact that the field equations of~(\ref{5daction}) with $(g, \cF)$ invariant under a $U(1)^2$ isometry group are equivalent to those of a 3d theory of gravity coupled to a non-linear sigma model with coset target space  $G_2 / SO(4)$ \cite{Cremmer}. This structure allows one to reduce the classification problem to a set of involved algebraic constraints. The results were recently reported on in~\cite{KL2011}.

\section{Summary}  Analyzing and classifying near-horizon geometries offers a systematic technique to map out the space of extremal black holes in four and higher dimensions.  In particular, we may deduce important properties of these black holes, such as the near-horizon $SO(2,1)$ symmetry enhancement, which hold in a generic class of gravitational theories. As the above examples demonstrate, the near-horizon approach is especially useful in the context of asymptotically AdS spacetimes because established solution-generating techniques are not available.  A key property of the near-horizon geometry is that one can extract useful information (e.g. geometry, topology, and some conserved charges) that is relevant for both the classification problem and for the quantum description of black holes within string theory and the AdS/CFT correspondence.  Clearly, there are a number of outstanding problems remaining to be solved regarding the near-horizon classification problem.  More ambitiously, it would be useful to determine the conditions necessary to evolve a near-horizon geometry outwards to a black hole with specified asymptotics. This is essential to strengthen these tools to establish uniqueness theorems for supersymmetric and extremal black holes. \\

\noindent {\bf Acknowledgements}: This research is supported by a fellowship from the Pacific Institute of the Mathematical Sciences and NSERC. The author wishes to particularly thank Harvey Reall and James Lucietti, with whom much of this work was first completed, as well as the scientific organizing committee of GR-19.

\end{document}